\def\year{2020}\relax
\documentclass[letterpaper]{article} 
\usepackage{aaai20}  
\usepackage{times}  
\usepackage{helvet} 
\usepackage{courier}  
\usepackage[hyphens]{url}  
\usepackage{graphicx} 
\usepackage{graphicx}  
\usepackage{multirow}

\title{Seeing through the smoke : a world-wide comparative study of e-cigarette flavors, brands and markets using data from reddit and twitter}
\author{\Large \textbf{Rohit Venkata Sai Dulam, \textsuperscript{\rm 1} Meghana Murthy, \textsuperscript{\rm 2}Jiebo Luo\textsuperscript{\rm 3}}\\ 
\textsuperscript{\rm}University of Rochester\\ 
rdulam@ur.rochester.edu, \textsuperscript{\rm 1} mmurthy@ur.rochester.edu, \textsuperscript{\rm 2} jluo@cs.rochester.edu \textsuperscript{\rm 3}\\ 
}
 
\begin{document}

\maketitle

\begin{abstract}
The growing popularity of E-cigarettes, an alternative to cigarettes, have motivated us to study trends pertaining to the brands, flavors and online market activity using posts from Reddit and Twitter. The main motivation for this world-wide study is to emphasize the difference that laws and regulations have on the usage and availability of different flavors and brands of vapes in different countries. Data has been obtained from subreddits belonging to e-cigarette communities from Australia, Canada, Europe and the UK. Extensive cleaning of data, and rigorous text mining operations provide varying results for different countries. We expect the results obtained from Twitter and Reddit to be different, given the different atmospheres they provide to users. Each tweet allows 280 characters for users whose identities are known, whereas Reddit posts are limited to 40,000 characters and user identities remain anonymous. In the future, this work can be extended to posts apart from the English language.   
\end{abstract}

\section{Introduction}
The growing importance of combating usage of e-cigarettes is evident by the increasing death tolls in the USA. A 2018 National Academy of Medicine report discovered that young people who use e-cigarettes may be more likely to smoke cigarettes in the future. As of November 21st 2019, CNBC reports vaping lung disease case total rises to 2290, while the death toll reaches 47. The reason for hospitalization has been termed as EVALI by medical professionals - E-cigarette, or Vaping, product use Associated Lung Injury. These findings make it paramount to understand the market for e-cigarettes in various countries which in turn can help the concerned authorities on taking necessary steps to curb it's usage. 

Previous work has mostly focused on the USA. As different countries have different laws and regulations related to nicotine and e-cigarettes, it is important to go beyond the scope of America. For example, Australia has banned products containing nicotine. The UK, on the other hand, has not banned nicotine, but limits the strength of e-liquids and its container size. Japan has banned nicotine altogether, but in Canada very few regulations exist - it is completely legal for personal use.  This difference between countries makes it important to understand that the results obtained in each case need not be the same, considering the laws and regulations pertaining to e-cigarettes aren’t the same either. 

In America, JUUL is currently the top-selling e-cigarette brand. Other examples include the MarkTen Elite, Aspire, Kanger and the PAX Era. They are available in different flavors including fruit, candy and creme.  So far, there has been an increasing interest in finding a relation between vape flavors and their respective health issues Li et al.

For this project, the focus is on Canada, Australia, the United Kingdom and countries in Europe. Data collected from Reddit and Twitter provides information from various vape users writing posts about e-cigarette brands, their prices and their popularity. With this information, online markets can be found. These online markets, grouped by location, provide information on which states in countries have the highest e-cigarette online purchasing activity. 

Due to the anonymity of Reddit, there are a countless number of personal vaping experiences shared by users. Hence, it is possible to study patterns and trends JP et al related to brands and flavors. It would also be interesting to study the user profiles of e-cigarette users. However, unless explicitly mentioned, anonymity makes it hard for us to determine the age and gender, as both remain unknown. For this reason, specific subreddits that users subscribe to or follow have to be analysed to provide an estimate of what age group they belong to.

\begin{table}[h!]
\caption{Subreddit Communities}\smallskip
\centering
\smallskip\begin{tabular}{l}
\hline
r/ecigclassifieds, European Vapers, r/ecr\_eu, \\ r/Teenagers,  r/Canadian\_ecigarette, r/aussivapers \\
\hline
\end{tabular}
\label{table1}
\end{table}

 Table 1 shows the subreddit communities used to mine information. 

In this way, we perform a comparative study between different countries focusing on the brands, prices and markets of e-cigarettes. The main contributions of our research work includes: 
\begin{enumerate}
    \item We analyze Reddit posts and comments pertaining to e-cigarette communities of specific countries to better understand its usage and the presence of online markets. 
    \item We determine the popularity of flavors and brands of e-cigarettes by studying streamed tweets.
    \item We analyze temporal trends of popularity of brands in different countries based on Reddit posts and upvotes of comments. 
\end{enumerate}

\section{Related Work}
The concept of studying alcohol and drug usage through social media(Luo et al. 2016) shows how user demographics are studied to observe a trend between the time and location of Instagram posts. Additionally, the age and gender of the Instagram user is considered. Accounts followed by these users are also studied to obtain a common interest these users may share. This motivated us to perform an extensive analysis on Reddit posts and comments, to examine which flavors and brands e-cigarette users might be interested in. 

A study conducted by Majmundar A et al.(2019) provides us with an insight of the serious health risks associated with second-hand and third-hand smoke. These insights can further be used to strengthen policies related to public e-cigarette usage and strengthen nicotine and tobacco regulations. They collected relevant tweets using Twitter’s Streaming API and filtered the tweets by using 26 vape related terms. This study was limited by collected data from only the year of 2018. Also, the study cannot be generalized to other social media platforms, Twitter was the only platform in consideration. This motivates us to choose 2 social media platforms - Reddit and Twitter. 

Sharma et al. (2016) discusses the importance of online discussions about e-cigarettes and mental health. Data is collected from Reddit which is freely accessible in the public domain. It was deduced that the motivations for using e-cigarettes vary between self-medication, to quit smoking, as a hobby, and to feel a sense of social connectedness. 

Dai, H., \& Hao, J. (2017) study also uses Twitter as a source of data by collected e-cigarette related discussions. It also emphasizes on the importance of realizing the difference between human users and social bots while using social media to collect public-health related data. Opinion polarities motivates us to attempt a sentiment analysis, or obtain a metric of popularity of e-cigarette related laws, brands and flavors by measuring favorite counts, retweets and upvotes on Twitter and Reddit.  

In a different study,data was collected from all over the world between the years 2011 and 2015(Wang L et al. 2015). Popularity of flavors is measured by counting the number of times each flavor was mentioned in a post.  We also measure popularity in a similar manner, however, we will not consider the entire dataset all at once. A more thorough analysis can be performed by grouping data based country-wise.

(Barker et al. 2019) talks obtaining and clustering posts obtained from reddit into particular categories. Also, they obtain tweets from across the whole Reddit for a particular period of time and do their analysis. Our work is primarily geared towards finding patterns in countries outside of the US and see how these communities have changed over time with the ongoing resistance to vaping products due to health problems caused by them.

\section{Data and Extracted Features}
\subsection{Twitter}
To collect data from Twitter, the Python library Tweepy is used to access the Twitter API. Streaming tweets with Tweepy provides a live-feed of tweets that are collected which can be filtered based on keywords specified by the user. Some keywords we used to filter tweets and obtain relevant data are given below. 

\begin{table}[h!]
\caption{Keywords}\smallskip
\centering
\smallskip\begin{tabular}{l}
\hline
vape,vaping,e-cig,JUUL,e-cigarette, \\electronic cigarette, vapestick,ehookah, e-juice, \\e-swisher,smoke assist, vaper, rta, rba, rdta\\
\hline
\end{tabular}
\label{table1}
\end{table}

The results are obtained in a JSON format which has been written to a .txt file. Using Pandas, an open-source data analysis tool offered by Python, we can read the file into a data frame to perform further analysis. 

Some major limitations of this method is the streaming limit. Twitter API allows a maximum of 3200 tweets for extraction per request, so the data collection takes a tremendous amount of time.

The tweets obtained contain numerous special characters, user names, retweets content, URLs and emojis. The cleaning process is extensive, making use of the ‘Emoji’ and ‘re’ packages offered by Python to simplify the process. Once the tweets are entirely cleaned, they are ready to be used for text mining to obtain useful information.    

\subsection{Reddit}
Data from Reddit is collected using the PushShift and Praw API. Praw is the Python API Reddit wrapper and is used to access Reddit data.  Like any other text retrieval process, the data at the first step is not entirely clean. The cleaning process includes removing special characters and stop words, lemmatization and some basic text mining procedures. \\

\begin{figure}[h]
\begin{center}
\includegraphics[width=1.0\columnwidth]{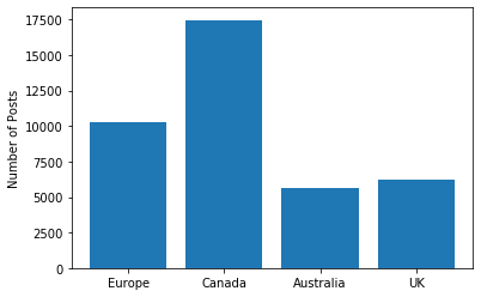}
\end{center}
\caption{Number of posts collected from different subreddits.}
\end{figure}

Subreddits specific to Europe, Canada, Australia and the United Kingdom were chosen in order to fully understand e-cigarette users specific to their own country. The Canadian subreddit had the highest number of users. Rest of the subreddit communities had approximately the same number of users.

From figure 1, it is evident that Canada has the post number of posts related to e-cigarettes. This could be attributed to more active members in the subreddit, and fewer amount of laws and regulations pertaining to e-cigarettes, thereby freely allowing e-cigarette users to explore more brands and flavors and further post about their experiences online.

\section{Methodology}
\subsection{Data Collection}
As mentioned in the Data section of this paper, the data is collected from Twitter and Reddit using their respective API’s. Data from Reddit includes both the posts and their associated comments. Since comments also contain significant information, we felt like it is important to utilize it as well. 
\subsection{Data Cleaning}
The data is cleaned and manipulated per our requirements by extensively using the Pandas library on Python.
\subsection{Flavors}
\begin{table}[h!]
\caption{Flavors}\smallskip
\centering
\smallskip\begin{tabular}{l}
\hline
Watermelon, Spearmint, Banana, Vanilla, Strawberry\\, Peanut Butter, Butterscotch, Pineapple, Apple, \\Blueberry, Cinnamon Roll, Pomegranate, Caramel, \\Menthol, Raspberry\\
\hline
\end{tabular}
\label{table1}
\end{table}
The flavors have been handpicked after reading numerous articles online. Once we shortlisted 10-15 flavors indicated in Table 3, popularity of these flavors were determined from Reddit posts. Here, popularity is defined by the number of tweets or posts the specific flavor has been mentioned in. 
\subsection{Brands}
First, we used the BeautifulSoup package offered by Python to scrape a web page by the name of ‘misthub’ to get the names of all companies that manufacture e-cigarettes. Next, we determined the top 5 most common brands based on their number of occurrences on Reddit. These brands include Aspire, Kanger, Eleaf, One Up Vapor and Innokin. Surprisingly, Juul - despite being very popular in the USA among teenagers and young adults, was not found to be too common in other countries on Reddit. However, we have included it in the study to analyse this difference in popularity. 
\subsection{Online Markets}
The “E-Cig Classifieds” subreddit is an online community where users buy and sell vapes. Analysing data from this subreddit helps gain a better insight of which areas have the highest online buying/selling activity when compared to others. They have a unique way to define the title of posts. Any user who buys or sells vaping products has to specify their location i.e. particular state in the US etc. This information was very helpful in determining locations with high vaping activity.

\section{Experiments and Discussions}
\subsection{Data Analysis of Online Markets - Reddit}
From Figure 2,in the USA, California, New York, Texas, Mississippi, Florida, Washington and Kentucky are some prominent states. Interestingly, according to a September 2019 article by the USNews, Kentucky is the \#2 state in America having the highest e-cigarette consumption.
\begin{figure}[h]
\begin{center}
\includegraphics[width=1.0\columnwidth]{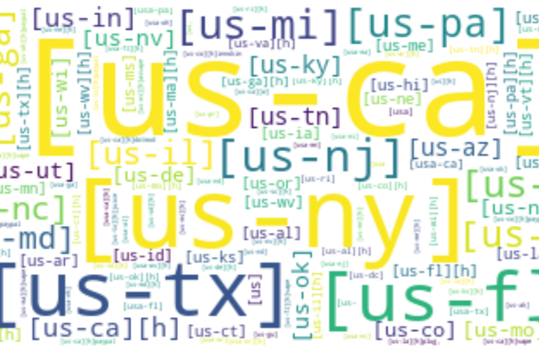}
\end{center}
\caption{Word Cloud of States in the US based on the strength of online e-cigarette markets present.}
\end{figure}

\begin{figure}[h]
\begin{center}
\includegraphics[width=1.0\columnwidth]{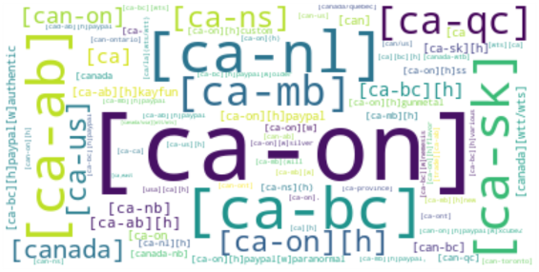}
\end{center}
\caption{Word Cloud of Provinces in Canada based on the strength of online e-cigarette markets present.}
\end{figure}

Figure 3 shows provinces in Canada having the highest buying/selling online activity include Ontario, British Columbia, Alberta, and Quebec, to name a few.

Figure 4 shows that in Europe (including the UK), the United Kingdom dominates the online vape market presence, followed by Germany, France, Norway, Netherlands, et cetera.

\begin{figure}[h]
\begin{center}
\includegraphics[width=1.0\columnwidth]{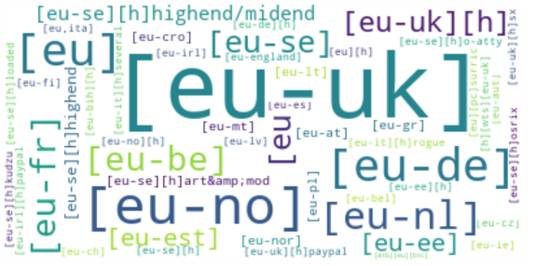}
\end{center}
\caption{Word Cloud of Countries in Europe based on the strength of online e-cigarette markets present.}
\end{figure}

The “ECR\_EU” subreddit is a general discussion page that e-cigarette users from the European Union use. To classify the subreddit posts into their respective countries, the Pycountry Python library was used. Figure 5 illustrates these findings.

\begin{figure}[h]
\begin{center}
\includegraphics[width=1.0\columnwidth]{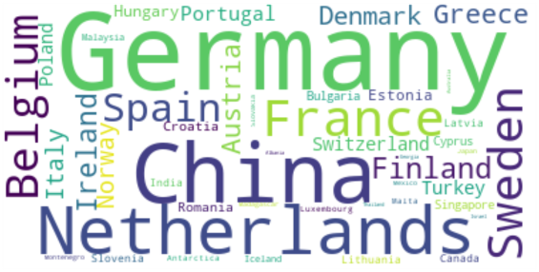}
\end{center}
\caption{Word Cloud of Countries having highest presence in the “ECR\_EU” subreddit.}
\end{figure}

Interestingly, China is mentioned numerous times in a subreddit pertaining to the EU. This can be attributed to the fact that products associated with China have lower cost of manufacturing and shipping combined. 
\subsection{Data Analysis of Flavors - Twitter}
Of all the tweets collected, 16,433 are in English. Out of all these English tweets, only 876 contain the word ‘flavor’ or ‘flavour’. However, analyzing the data for occurrences of specific flavors, we can get an estimated popularity of each flavor. 

From figure 7 it is clear that Menthol has the highest popularity, followed by apple, strawberry and banana. Due to the recent ban of Menthol vapes in America, we would also expect Menthol to be the most spoken about on social media platforms. 
\subsection{Data Analysis of Brands - Twitter}
Based on the data collected from Twitter, JUUL is the only brand being spoken about. A few other brands like ‘Vaporesso’ and ‘Vandy vape’ are mentioned a few times, but these instances are insignificant when compared to the presence of JUUL. Figure 8 shows these findings.

The high presence of JUUL on Twitter can also be attributed to the recent laws regarding JUUL flavor bans. It has been among the most common vape brand in America since the end of 2017 and it’s extensive use by youth has prompted concern from various public health communities. 

\subsection{Popularity Variance of Brands Over Time - Reddit}

We notice the increase or decrease in popularity of brands based on how often or not they are being spoken about on Reddit posts over the years. One limitation in counting the number of occurrences of each brand name is the Pandas string comparison operation. For example, the e-cigarette brand “alternativ” is a substring of the word “alternative”. Pandas counts “alternative” as an occurrence of the brand name, which causes an error in the results. To overcome this, we use regular expression comparisons instead to get an accurate count.

\begin{figure}[h!]
\begin{center}
\includegraphics[width=1.0\columnwidth]{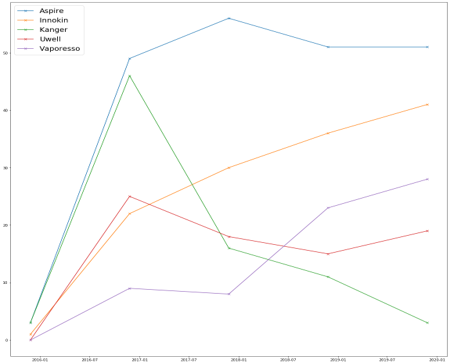}
\end{center}
\caption{Popularity Variance of Brands Over Time in the UK subreddit.}
\end{figure}

In UK, the trends are slightly different. They mostly resemble an increase, decrease, followed by a slight increase again. These differences in trends is evident of the difference in severeness of laws and regulations pertaining to the production, advertising and selling of e-cigarettes across the globe.

\begin{figure}[h!]
\begin{center}
\includegraphics[width=1.0\columnwidth]{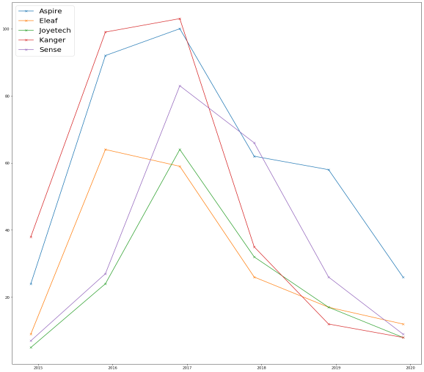}
\end{center}
\caption{Popularity Variance of Brands Over Time in the Canadian subreddit.}
\end{figure}
Canada, just like Europe, has consistent decreasing trend lines in the popularity of brands over time. This is interesting, given the relatively lenient laws in place. Perhaps the decrease is related to the growing number of health issues and lung diseases that can be contracted.

\begin{figure}[h!]
\begin{center}
\includegraphics[width=1.0\columnwidth]{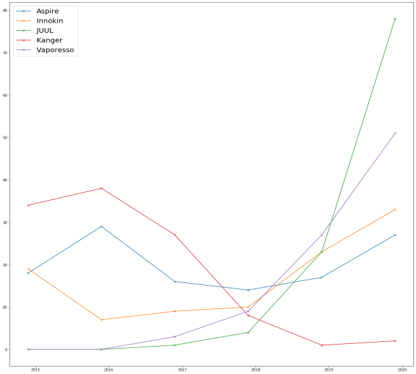}
\end{center}
\caption{Popularity Variance of Brands Over Time in the Australian subreddit.}
\end{figure}

Australia is the only country in this study with JUUL being in the top 5 most popular brand (apart from the USA). Australia, slightly similar to UK, shows an initial decrease and final increase in brand popularity on Reddit.

\begin{figure}[h!]
\begin{center}
\includegraphics[width=1.0\columnwidth]{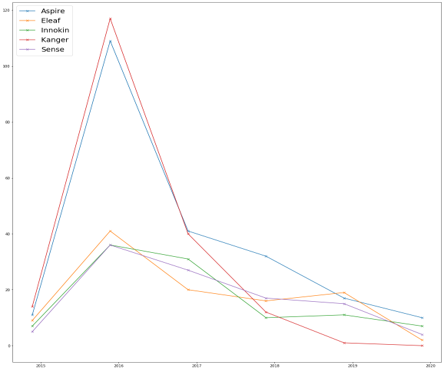}
\end{center}
\caption{Popularity Variance of Brands Over Time in the European subreddit.}
\end{figure}

EUROPE
In Europe, we notice a general decreasing trend in brand popularity.

\subsection{Popularity of Flavors across Countries - Reddit Posts}
The popularity of flavors across different countries is analysed by comparing the number of occurrences in different subreddits specific to different countries. In order to plot the Word Cloud, all the occurrences of a particular flavor have been summed up. From figure 10, it is evident that menthol, strawberry, banana, Apple and Vanilla seem to be among the popular flavors. The popularity of brands in individual subreddits comes out very similar to the aggregated word cloud, hence making it a perfect illustration. 

\begin{figure}[h]
\begin{center}
\includegraphics[width=1.0\columnwidth]{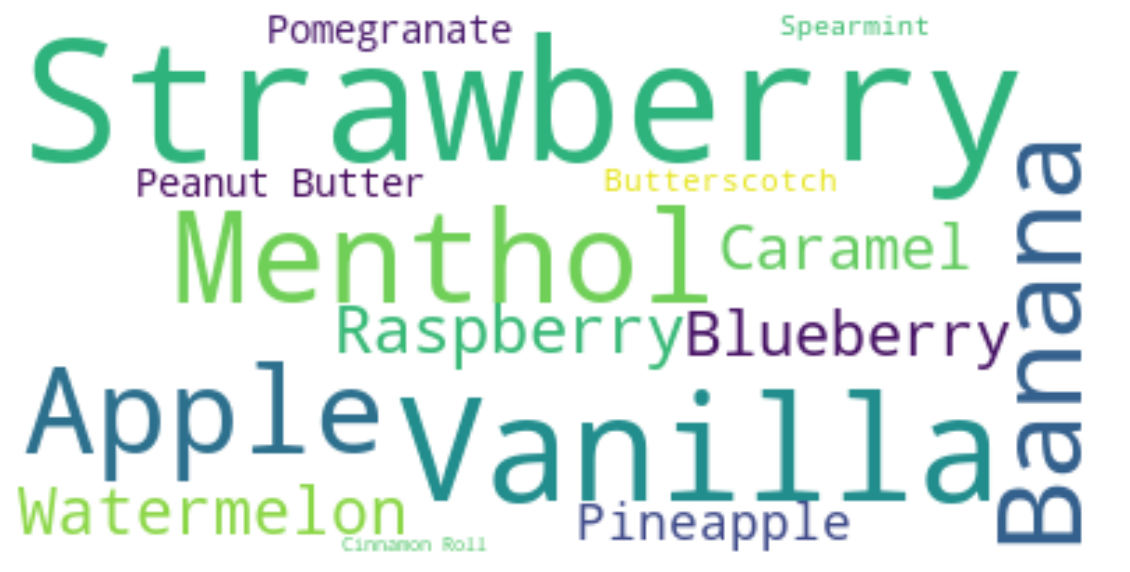}
\end{center}
\caption{Word Cloud of popular flavors across different subreddits.}
\end{figure}

\subsection{Data Analysis within America - Reddit}
\begin{table}[h!]
\caption{List of subreddits used to analyze the market in the United States}\smallskip
\centering
\smallskip\begin{tabular}{l}
\hline
BostonVapers, Chicago\_Vapers,Cleveland\_Vapers,\\ColoradoVapers,
Columbus\_Vapers,FL\_Vapers,HawaiiVapers,\\Maryland\_Vapers,MittenVapers,
MNvapers,Missouri Vapers,\\NH\_Vapers,NJ\_Vaping,NorthernNevadaVapers,
NYvapers,\\OK\_Vapers,orangecountyvapors,Philly\_Vapers,pghvape,
\\RVApers,BayAreaVapers,Santa\_Clarita\_Vapers,scvapes,\\SoCalVapers,
SouthEasternVapers,Texas\_Vaping,\\Utah\_Vapers,VapingInPhoenix,
TNVaporEnthusiasts,\\WashingtonStateVapers,WNYvape\\
\hline
\end{tabular}
\label{table1}
\end{table}
So far, the study focuses on countries outside of the US as plenty of work has already been done pertaining to within America. However, our analysis of e-cigarette users within America through Reddit data did not amount to much as sufficient information was not available. There wasn't a single subreddit that pertained to the US E-cigarette community, resulting in very small amounts of data. Interestingly, every state in the US had its own subreddit for e-cigarettes. They are listed in Table 4. We used the posts from these subreddits to plot the word cloud i.e. Figure 10. The posts were inadequate and hence no particular trend could be found out. 
Thankfully, we can rely on Twitter as a source of large number of tweets pertaining to e-cigarettes, at the cost of losing anonymity and a larger number of characters allowed in posts.

\begin{figure}[h!]
\begin{center}
\includegraphics[width=1.0\columnwidth]{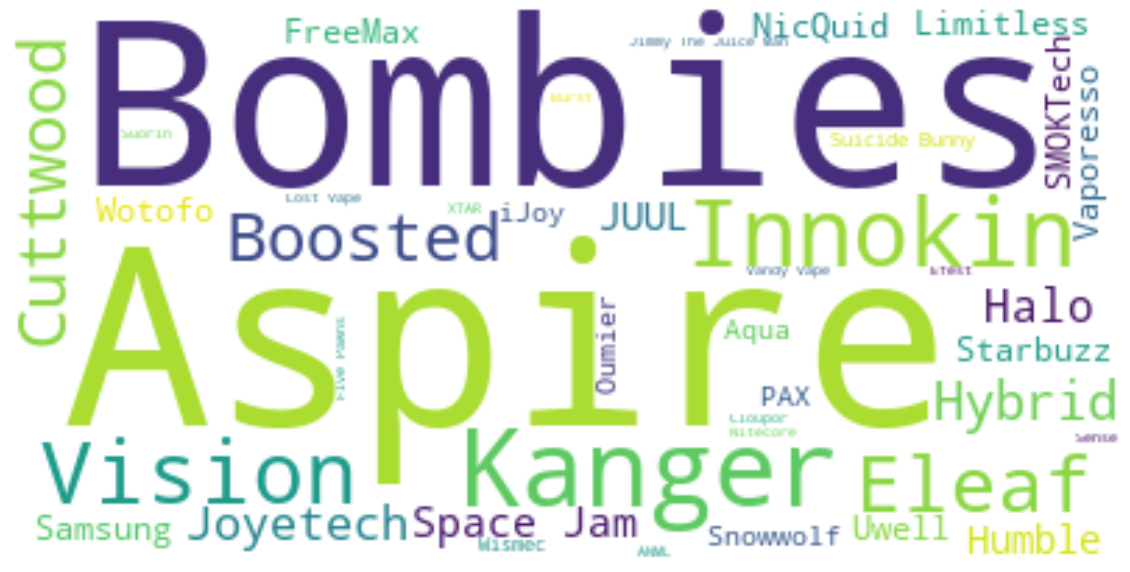}
\end{center}
\caption{Word Cloud of Popular brands obtained from US subreddits.}
\end{figure}

\section{Conclusion}
Vaping has largely been considered an alternative for cigarettes, which leads us to believe the increasing trends in initial years comes from an effort put in by smokers to combat their addiction. Our project is a comparative study of user demographics in countries outside of the US popular for vaping, and we obtain different results from each country/continent.Overall, our results provide reasonable proof of the effect laws and regulations pertaining to different countries have on the flavors, brands and online markets in the realm of e-cigarettes.

\section{Future Work}
As the study is world-wide, it is important to obtain data outside of the English language. Tweets from other languages have been collected, like Japanese, Spanish, French and Indonesian. In the future, we can translate these tweets and observe if more trends are hidden in foreign-language social media posts. 

Additionally, certain Twitter pages likes ‘ParentsvsVape’ and ‘VapeRite’ tweet extensively about laws, opinions, and the current happenings about e-cigarettes. These specific pages can be analysed, instead of collected tweets filtered through keywords. 

Finally, a more in-depth analysis of the information obtained through the comments on individual posts can also result in important trends. Our work has ended with finding temporal patterns of different brands and flavors, the next step would be to find the effectiveness of each brand and what is the sentiment associated with the posts and their respective comments.

\section{References}

\noindent Anuja Majumdar, Jon-Patrick Allem, Tess Boley Cruz and Jennifer B. Unger, “Where Do People Vape? Insights from Twitter Data”  \textit{International Journal of Environmental Research and Public Health.} 

\noindent Ratika Sharma, Britta Wigginton, Carla Meurk, Pauline Ford and Coral E. Gartner, “Motivations and Limitations Associated with Vaping among People with Mental Illness: A qualitative Analysis of Reddit Discussions”, \textit{International Journal of Environmental Research and Public Health} 

\noindent van der Tempel, J., Noormohamed, A., Schwartz, R. et al., Vape, quit, tweet? Electronic cigarettes and smoking cessation on Twitter. \textit{Int J Public Health (2016),} 61:249

\noindent Dai H, Hao J, Mining social media data for opinion polarities about electronic cigarettes  \textit{Tobacco Control 2017}. 175-180

\noindent Li, Q., Zhan, Y., Wang, L. et al. Analysis of symptoms and their potential associations with e-liquids’ components: a social media study. \textit{BMC Public Health (2016)} 16,674

\noindent Allem JP, Ferrara E, Uppu SP, Cruz TB, Unger JB, E-Cigarette Surveillance With Social Media Data: Social Bots, Emerging Topics, and Trends,  \textit{JMIR Public Health Surveill 2017:} 3(4):e98

\noindent Joshua O. Barker, MS1 , and Jacob A. Rohde, MA, Topic Clustering of E-Cigarette Submissions Among Reddit Communities: A Network Perspective

\noindent Yiheng Zhou, Numair Sani, and Jiebo Luo, "Fine-grained Mining of Illicit Drug Use Patterns Using Social Multimedia Data from Instagram," Special Session on Intelligent Data Mining,"  \textit{IEEE International Conference on Big Data (Big Data),} Washington, DC, December 2016.

\noindent Yiheng Zhou, Jingyao Zhan and Jiebo Luo, "Predicting Multiple Risky Behaviors via Multimedia Content," \textit{International Conference on Social Informatics (SocInfo)} ,Oxford, England, September 2017.

\noindent Xitong Yang, Jiebo Luo, "Tracking Illicit Drug Dealing and Abuse on Instagram using Multimodal Analysis," \textit{ACM Transactions on Intelligent Systems and Technology,}8(4): 58:1-58:15, August 2017.

\noindent Ran Pang, Agustin Baretto, Henry Kautz, and Jiebo Luo, "Monitoring Adolescent Alcohol Use via Multimodal Data Analysis in Social Multimedia," Special Session on Intelligent Mining, \textit{IEEE Big Data Conference,}Santa Clara, CA, October 2015.

\noindent Jianbo Yuan, Quanzeng You, and Jiebo Luo, "Are There Cultural Differences in Event Driven Information Propagation Over Social Media?" \textit{ACM Multimedia Conference, International Workshop on Socially-Aware Multimedia (IWSAM),} October 2013

\noindent Yiheng Zhou, Numair Sani, Jiebo Luo, "Fine-grained mining of illicit drug use patterns using social multimedia data from instagram", \textit{Fine-grained mining of illicit drug use patterns using social multimedia data from instagram} December 2016

\noindent Joshua O. Barker, Jacob A. Rohde, "Topic Clustering of E-Cigarette Submissions Among Reddit Communities: A Network Perspective",
\textit{} December 2019

\end{document}